\documentclass[%
 reprint,
superscriptaddress,
nofootinbib,
 amsmath,amssymb,
 aps,
]{revtex4-2}

\usepackage{graphicx}% Include figure files
\usepackage{dcolumn}% Align table columns on decimal point

\usepackage{braket, amsthm}

\usepackage[colorlinks,citecolor=blue,linkcolor=blue, urlcolor=blue, anchorcolor=blue]{hyperref}
\usepackage[ruled]{algorithm2e}

\newtheorem{lemma}{Lemma}
\newtheorem{theorem}{Theorem}
\newtheorem{corollary}{Corollary}
\newtheorem{definition}{Definition}

\theoremstyle{definition}

\begin{document}

\preprint{APS/123-QED}

\title{Deterministic quantum  search on all Laplacian integral graphs}% Force line breaks with \\

\author{Guanzhong Li}
\thanks{co-first authors}
% \email{ligzh9@mail2.sysu.edu.cn}
\affiliation{Institute of Quantum Computing and and Software, School of Computer Science and Engineering, Sun Yat-sen University, Guangzhou 510006, China}
\author{Jingquan Luo}
% \email{luojq25@mail2.sysu.edu.cn}
\thanks{co-first authors}%These authors contributed equally to this work.
\affiliation{Institute of Quantum Computing and and Software, School of Computer Science and Engineering, Sun Yat-sen University, Guangzhou 510006, China}

\author{Shiguang Feng}
%\email{fengshg3@mail.sysu.edu.cn}
\affiliation{Institute of Quantum Computing and and Software, School of Computer Science and Engineering, Sun Yat-sen University, Guangzhou 510006, China}

\author{Lvzhou Li}
\email{lilvzh@mail.sysu.edu.cn}
% \altaffiliation[Also at ]{Quantum Science Center of Guangdong-Hong Kong-Macao Greater Bay Area, Shenzhen, 518045, China}
\affiliation{Institute of Quantum Computing and and Software, School of Computer Science and Engineering, Sun Yat-sen University, Guangzhou 510006, China}
\affiliation{Quantum Science Center of Guangdong-Hong Kong-Macao Greater Bay Area, Shenzhen, 518045, China}

\date{\today}% It is always \today, today,

\begin{abstract}
Searching for an unknown marked vertex on a given graph (also known as spatial search) is an extensively discussed topic in the area of   quantum  algorithms, with a plethora of results based on different quantum walk models and targeting various types of graphs.
Most of these algorithms have a non-zero probability of failure. In recent years, there have been some efforts to design quantum spatial search algorithms with $100\%$ success probability. However, these works either only work for very special graphs or only for the case where there is only one marked vertex.
In this work, we  propose a different and  elegant approach to quantum spatial search, obtaining  deterministic quantum search algorithms that can find a marked vertex with certainty on any Laplacian integral graph with any predetermined proportion of marked vertices. Thus, this work discovers the largest class of graphs  so far that allow deterministic quantum search, making it easy  to design deterministic quantum search algorithms for many graphs, including the different graphs discussed in previous works, in a unified framework.

%Recently, Wang et al. [Phys. Rev. A 111, 042608 (2025)] proposed a unified quantum spatial search algorithm which can find a marked vertex with  probability $100\%$ on a large class of  Laplacian integral graphs. However, the  shortcomings of the algorithm are obvious: it has high time complexities for certain graphs, only works for the case of a single marked vertex, and requires additionally the graph to be vertex transitive.
%Our algorithm is based on a recently proposed model named controlled intermittent quantum walks, which enables us  to design a more universal  deterministic spatial search algorithm.
% Our result improves the result by Wang et al. [Phys. Rev. A 111, 042608 (2025)], as their algorithm has worse complexities in certain graphs, only works for the case of a single marked vertex, and requires additionally the graph to be vertex transitive.
%Our algorithm is based on a recently proposed model named controlled intermittent quantum walks, which is a simple generalization of the alternating quantum walks by allowing the walk operation to be controlled.
%Interestingly, with this slight modification of the model, we are able to design a more universal  deterministic spatial search algorithm using a different and more elegant approach.
\end{abstract}


\maketitle

%\tableofcontents

% \begin{table*}
% \caption{\label{tab:table3}This is a wide table that spans the full page
% width in a two-column layout. It is formatted using the
% \texttt{table*} environment. It also demonstates the use of
% \textbackslash\texttt{multicolumn} in rows with entries that span
% more than one column.}
% \begin{ruledtabular}
% \begin{tabular}{ccccc}
%  &\multicolumn{2}{c}{$D_{4h}^1$}&\multicolumn{2}{c}{$D_{4h}^5$}\\
%  Ion&1st alternative&2nd alternative&lst alternative
% &2nd alternative\\ \hline
%  K&$(2e)+(2f)$&$(4i)$ &$(2c)+(2d)$&$(4f)$ \\
%  Mn&$(2g)$\footnote{The $z$ parameter of these positions is $z\sim\frac{1}{4}$.}
%  &$(a)+(b)+(c)+(d)$&$(4e)$&$(2a)+(2b)$\\
%  Cl&$(a)+(b)+(c)+(d)$&$(2g)$\footnotemark[1]
%  &$(4e)^{\text{a}}$\\
%  He&$(8r)^{\text{a}}$&$(4j)^{\text{a}}$&$(4g)^{\text{a}}$\\
%  Ag& &$(4k)^{\text{a}}$& &$(4h)^{\text{a}}$\\
% \end{tabular}
% \end{ruledtabular}
% \end{table*}

\section{Introduction}
Quantum walk, the analogy to classical random walk, is a widely adopted paradigm to design quantum algorithms for various problems, such as spatial search~\cite{coins_05, CG_04, quadratic_20, quadratic_22, unified, PhysRevA.106.052207, universal}, element distinctness~\cite{element_ambainis}, matrix product verification~\cite{matrix_product}, triangle finding~\cite{MagniezSS07}, group commutativity~\cite{MagniezN07}, the welded-tree problem~\cite{CCD03}, the hidden flat of centers problem~\cite{nonlinear}, and so on.
% ,edp_2024
% ,multi2023,recovering, multi_electric, global_phase
There are two types of quantum walks: the continuous-time quantum walk (CTQW) and the discrete-time quantum walk (DTQW).
CTQW is relatively simple, and mainly involves simulating a Hamiltonian $H$ that encodes the structure of the graph.
DTQW is more diverse, ranging from the earliest and simplest coined quantum walk~\cite{AmbainisBNVW01,AmbainisKV01} to various Markov chain based frameworks~\cite{Szegedy_03,MNRS,belovs2013quantum,KroviMOR16,unified}.

As one of the most important algorithmic applications of quantum walks, spatial search is the fundamental problem of finding a marked vertex on a given graph.
Quantum algorithms for spatial search on generic graphs have been designed via DTQW~\cite{quadratic_20,unified} and CTQW~\cite{quadratic_22}, providing a general quadratic speedup compared with classical search algorithms.
Besides these elegant results, a large body of research on quantum spatial search algorithms focuses on specific graphs~\cite{coins_05, CG_04, hypercube_09, Janmark2014, bipartite_19, RN3, RN4, johnson_discrete, HG, Chakraborty2016, PhysRevA.106.052207}, aiming at designing a quantum algorithm with $O(\sqrt{N})$ cost for a given graph with $N$ vertices.
However, all these algorithms are probabilistic and have a non-zero probability of failure.

Deterministic spatial search on graphs, i.e. finding a marked vertex with probability $100\%$, has been less studied until recently.
Deterministic quantum search algorithms were proposed for complete bipartite graphs based on CTQW~\cite{bipartite_24} and DTQW~\cite{PRR_bipartite_24}, and for  complete identity interdependent networks (CIINs)~\cite{alternating_PRA} and star graphs~\cite{alternating_PRL} based on alternating quantum walks~\cite{alternating_QST}.
The significance of designing deterministic quantum algorithms lies not only in theoretically proving that the success probability can essentially be $100\%$ without sacrificing quantum speedup, but also in experimentally improving the success rate when the algorithm is implemented on a real quantum computer.

The research on designing deterministic quantum algorithms  can be referred to as derandomization of quantum algorithms, that is, the  process of turning probabilistic quantum algorithms into deterministic ones while maintaining the original speedup.
In addition to deterministic spatial search on graphs, there are also some other quantum algorithms that have been  derandomized, such as deterministic quantum algorithms for Grover  search~\cite{amplitude_amplification,arbi_phase,Long,FXR},  Simon's problem~\cite{exact_simon} (and its generalization~\cite{GSP}), the element distinctness problem~\cite{edp_2024}, triangle finding~\cite{triangle_2023} and the welded tree problem~\cite{recovering}.

In an impressive recent work by Wang et al.~\cite{universal}, a novel and universal approach was proposed for designing deterministic quantum search algorithms on various Laplacian integral graph\footnote{The spectrum of the graph's Laplacian matrix consists entirely of integers.} that is also vertex-transitive, such as (see Appendix~\ref{subsec:integral} for their definitions): Johnson graphs, Hamming graphs, Kneser graphs, Grassmann graphs, Rook graphs with CIIN being a special case, Complete-square graphs and star graphs.
Their algorithm also works for complete bipartite graphs which are neither Laplacian integral nor vertex-transitive, and thus it unifies all the previous results on deterministic spatial search and provides new results on more graphs.

Despite being a general result, the deterministic quantum search algorithm by Wang et al.~\cite{universal} based on alternating quantum walks has some shortcomings:
\begin{enumerate}
    \item The algorithm's complexity is $O(2^{d_L}\sqrt{N})$ (see Theorem~\ref{thm:s_m}) and it is related to the depth $d_L$ of the graph Laplacian (see Definition~\ref{def:d_L}), which may depend on $N$, as we will see later on some graphs.
    Thus, the complexity can be greater than $O(\sqrt{N})$ and therefore not optimal in certain cases.
    % not only related to the number of vertices $N$, but
    % Thus, some parameters of the graphs are fixed (e.g. $k$ in the Johnson graph $J(n,k)$) to guarantee that $d_L$ is a constant.
    \item Only one marked vertex is allowed on the graph, and the algorithm fails when dealing with multiple marked vertices.
    \item The graph has to be vertex-transitive, which limits the applicability of the algorithm.
    We will later present an interesting type of Laplacian integral graphs that are not vertex-transitive and $2^{d_L}=O(N)$.
    % so that an exhaustive search can be performed with constant overhead
    % or the types of different vertices is at most a constant, as seen by the complete bipartite graph~\cite[Theorem~6]{universal} which has only two types of vertices
\end{enumerate}

In this paper, we overcome all the above limitations by proposing a quantum search algorithm that can find a marked vertex with certainty on any Laplacian integral graphs with any number of marked vertices.
Our algorithm is based on the recently proposed controlled intermittent quantum walk (CIQW) model~\cite{QPD}, which generalizes alternating quantum walks~\cite{alternating_QST} by replacing its CTQW with controlled-CTQW (see Section~\ref{subsec:CIQW} for the definition of CIQW).
Specifically, we obtain the following result.
\begin{theorem}\label{thm:known}
    For any Laplacian integral graph, a CIQW-based quantum algorithm can find a marked vertex with certainty in total evolution time $O(\frac{1}{\sqrt{\varepsilon}})$ and query complexity $O(\frac{1}{\sqrt{\varepsilon}})$, where  $\varepsilon$ is the proportion of marked vertices known in advance.
\end{theorem}

\begin{figure*}[hbt]
	\centering
	\includegraphics[width=0.9\textwidth]{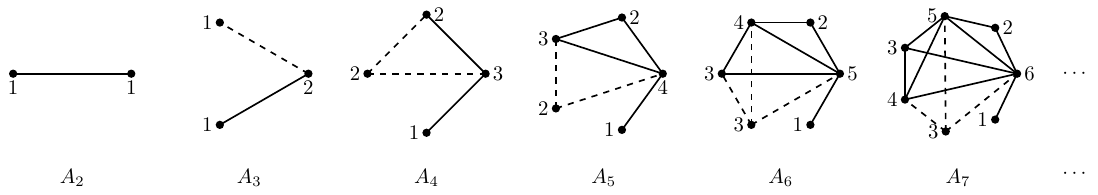}
    \caption{\label{fig:antiregular_graphs} Iterative construction of the first few antiregular connected graphs. The number beside each vertex denotes its degree. For $N\geq 2$, the graph $A_{N+1}$ can be obtained from $A_N$ by adding a new vertex, and connecting it to all vertices with degree greater than $\lfloor N/2\rfloor$ and to one of the two vertices with the same degree $\lfloor N/2\rfloor$, as shown by the dashed lines.}
\end{figure*}

To showcase the advantage of our result, we present three types of Laplacian integral graphs for which our algorithm  outperforms the algorithm given by Wang et al.~~\cite{universal}:
\begin{itemize}
    \item The antiregular connected graphs $A_N$ that have the following properties~\cite{ali2020survey}:
(i) $A_N$ are the only (under graph isomorphism) connected graphs with $N \geq 2$ vertices, such that the cardinality of the set of its vertex degrees is $N-1$ and the duplicated degree is $\lfloor N/2 \rfloor$;
(ii) $A_N$ can be constructed iteratively starting from $A_2=K_2$ as shown in Fig.~\ref{fig:antiregular_graphs}; 
and (iii) The Laplacian eigenvalues of $A_N$ are $\{0,1,\cdots,n\}\setminus\{\lfloor (N+1)/2 \rfloor \}$.
Thus, $A_N$ are not vertex-transitive by (i) so that their algorithm is not applicable, and the depth of $A_N$ are $d_L=O(\log(N))$ by (iii).
\item The Johnson graph $J(n,k)$ with $k=n^{2/3}$, which is used in the quantum walk based algorithm for element distinctness~\cite{element_ambainis}.
The Laplacian eigenvalues of it are $i(n+1-i)$ for $i\in\{0,1,\cdots,n^{2/3}\}$, and numerical simulation shows that $2^{d_L}=\Theta(n^{2/3})$.
\item The hypercube $H(n,2)$ that is the Hamming graph $H(d,q)$ with $d=n$ and $q=2$.
The Laplacian eigenvalues of it for $i\in\{0,1,\cdots,n\}$, and thus $d_L=\log(n)$.
\end{itemize}

We summarize the costs of our algorithm and the algorithm from~\cite{universal} on the three types of graphs in Table~\ref{tab:compare}.

\begin{table}[htb]
\caption{\label{tab:compare}Comparison of the costs of the previous algorithm and our algorithm on some specific graphs.}
\begin{ruledtabular}
\begin{tabular}{llll}
\textrm{Graphs}&
\textrm{Vertices}& %\footnote{Note a.}
\textrm{Wang et al.~\cite{universal}}& %\footnote{Note b.}
\textrm{Theorem~\ref{thm:known}} \\
% \colrule
\hline
$A_N$ & $N$ & -\footnote{The algorithm is not applicable for $A_N$, since $A_N$ are not vertex-transitive.} & $\sqrt{N/M}$\footnote{$M$ is the number of marked vertices, and the algorithm in ~\cite{universal} only works for $M=1$.} \\
$J(n,n^{2/3})$ & $N=\binom{n}{n^{2/3}}$ & $n^{2/3}\sqrt{N}$ & $\sqrt{N/M}$ \\
$H(n,2)$ & $N=2^n$ & $n\sqrt{N}$ & $\sqrt{N/M}$ \\
\end{tabular}
\end{ruledtabular}
\end{table}

\subsection{Related work}
In a recent work by Wang et al.~\cite{universal}, a succinct algorithmic framework via alternating quantum walks was presented, unifying quantum spatial search, state transfer, and uniform sampling on a large class of graphs.

The alternating quantum walks model~\cite{alternating_QST} can be seen as a mixtuire of CTQW and DTQW, and it interleaves CTQW $e^{iHt}$ with oracle queries $e^{i\theta \Pi_M}$, where $H$ encodes the structure of the graph and the Laplacian matrix or the adjacency matrix is commonly used, and $\Pi_M$ is the projection onto the subspace spanned by all the marked vertices.
We assume the initial state $\ket{\pi}$, i.e. the uniform superposition of all the vertices, is provided, and the system evolves as follows:
\begin{equation}
    \ket{\vec{t},\vec{\theta}} = \prod_{k=1}^{p} e^{iHt_k} e^{i\theta_k\Pi_M} \ket{\pi}.
\end{equation}
The goal is to choose the appropriate parameters $p,\vec{t},\vec{\theta}$ such that $\| \Pi_M \ket{\vec{t},\vec{\theta}} \|$ is close to 1.

Applying their framework to quantum spatial search, Wang et al.~\cite{universal} obtained the following result.
\begin{theorem}[Theorem~4 of Ref.~\cite{universal}]\label{thm:s_m} 
	Let $G=(V,E)$ be a vertex-transitive graph with $N$ vertices whose Laplacian matrix $L$ has only integer eigenvalues and $\ket{\pi}=\frac{1}{\sqrt{N}}\sum_{v\in V}  |v \rangle $.
    There is an integer $p \in O(2^{{d_L}}\sqrt{N})$ and real numbers $\gamma, \theta_j,t_j\in [0,2\pi)$   $(j\in \{1,2,\dots,p\})$, such that the following equation holds for all $m\in V$: \begin{align} \label{newtheq}
		\ket{\pi}=e^{-i\gamma}\prod_{j=1}^{p}e^{-i\theta_j |m\rangle\langle m|} e^{-iLt_j}|m\rangle.
	\end{align}		     
\end{theorem}
By reversing the above process, they obtained deterministic quantum spatial search on Laplacian integral graphs that are also vertex-transitive and have a single marked vertex $m\in V$.
The positive integer $d_L$ in Theorem~\ref{thm:s_m} is the depth of the Laplacian integral graph defined below, and it can be as large as the number of distinct eigenvalues of $L$ and as small as one.
\begin{definition}\label{def:d_L}
For a Laplacian integral graph, let $\Lambda_0 := \{\lambda_1,\dots,\lambda_N\}$ be its eigenvalues.
For $k\geq 0$, recursively define $\Lambda_{k+1}$ and $\overline{\Lambda}_{k+1}$ by
\begin{align}
    \Lambda_{k+1} &= \{\lambda\in\Lambda_ {k} : \frac{\lambda}{\gcd(\Lambda_ k)} \ \mathrm{even}\}, \\
    \overline{\Lambda}_{k+1} &= \{\lambda\in\Lambda_ {k} : \frac{\lambda}{\gcd(\Lambda_ k)} \ \mathrm{odd}\}.
\end{align}
Then $d_L$ is the least $k$ such that $\Lambda_k$ contains only $0$.
\end{definition}

This paper is organized as follows.
Section~\ref{sec:pre} introduces some background on spatial search and reviews the CIQW model.
We then present our algorithm in Section~\ref{sec:exact_search}, with an overview given in Section~\ref{subsec:overview} and a proof of the algorithm's correctness and the claimed complexity shown in Section~\ref{subsec:proof_thm}.
An analysis of the gate complexity is also discussed in Section~\ref{subsec:gate_complexity}.
Finally, Section~\ref{sec:summary} summarizes our work.

\section{Preliminaries}\label{sec:pre}
In this paper, we only consider simple undirected connected graphs, where ``simple” means the graph has no loops or multiple edges between any two vertices.
Let $G = (V,E)$ be a graph where $V$ is the vertex set and $E$ is the edge set.
The Laplacian matrix of $G$ is $L = D -A$, where $D$ is the diagonal matrix with $D_{jj} = \mathrm{deg}(j)$, the degree of vertex $j$, and $A$ is the $\{0,1\}$ adjacency matrix of $G$, where $A_{ij}=1$ if and only if $(i,j)\in E$.
In other words, the entries $L_{i,j}$ of $L$ is given by
\begin{equation}
    L_{i,j} =\begin{cases}
        -1, &\mathrm{if}\ i\neq j \ \mathrm{and}\ (i,j)\in E, \\
        0, &\mathrm{if}\ i\neq j \ \mathrm{and}\ (i,j)\notin E, \\
        \mathrm{deg}(j), &\mathrm{if}\ i= j.
    \end{cases}
\end{equation}

The Laplacian $L$ of a simple undirected graph $G$ is a symmetric matrix,  and  $L$ is diagonalizable.
Its spectrum has the following nice properties:
(i) All the eigenvalues of $L$ are non-negative and bounded above by $N := |V|$~\cite{Laplacian_distinct}.
(ii) $0$ is a simple eigenvalue (i.e. with multiplicity one) of $L$ if $G$ is connected, and the corresponding eigenvector is the uniform superposition of all vertices $\ket{\pi}:= \frac{1}{\sqrt{N}}\sum_{v\in V}\ket{v}$~\cite{Laplacian_spectral}.

\subsection{Spatial Search}

\begin{figure}[hbt]
    \includegraphics[width=0.45\textwidth]{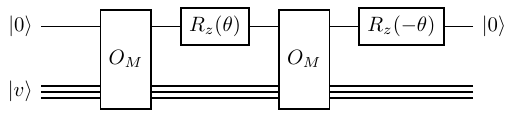}
	\caption{\label{fig:oracle} Quantum circuit implementation of the general oracle $e^{i\theta\Pi_M}$ (Eq.~\eqref{eq:general_oracle_def}) using the standard oracle $O_M$ (Eq.~\eqref{eq:standard_oracle}) and single qubit rotation $R_z(\theta) = \mathrm{diag}(e^{-i\theta/2},e^{i\theta/2})$. }
\end{figure}

Spatial search is the problem of finding an unknown marked vertex on a given graph $G$. When designing algorithms for this problem, it is generally assumed that there is an oracle checking whether a given vertex is the marked one, and the algorithm should invoke this oracle as few as possible.
In quantum computing, the standard oracle works as follows:
\begin{equation}\label{eq:standard_oracle}
    O_M\ket{b}\ket{v}=\ket{b \oplus f(v)}\ket{v},
\end{equation}
where $b \in \{0, 1\}$, $v \in V(G)$, and the Boolean function $f(v)=1$ iff $v$ is in the marked set $M$.
In this paper, we will use the general oracle $e^{i\theta \Pi_M}$ which works as: 
\begin{equation}\label{eq:general_oracle_def}
    e^{i\theta \Pi_M} \ket{v} =
    \begin{cases}
        e^{i\theta} \ket{v}, &{\rm if}\ v\in M; \\
        \ket{v}, &{\rm if}\ v\notin M.
    \end{cases}
\end{equation}
The general oracle can be implemented with two invocations of the standard oracle $O_M$ as shown in Fig.~\ref{fig:oracle}.

Quantum algorithms for spatial search are usually based on quantum walks.
The number of invocations of the oracle is defined as the query complexity of the algorithm. In addition, the circuit cost to implement the quantum walk is also concerned, which in general is proportional to the walk time. 
Thus, one should optimize the query complexity and the walk time when designing a quantum walk-based search algorithm.
%Therefore, in this paper, we aim to optimize the query complexity and the total evolution time of CTQW.
% \begin{remark}
%     One may want to account for the norm $\|L\|$ in the circuit cost to implement $e^{iLt}$.
%     We ignore this quantity as it is a constant once the graph under consideration is fixed.
% \end{remark}

%\note{maybe we should define the considered complexities more clear here.}

\subsection{Controlled intermittent quantum walks}\label{subsec:CIQW}
\begin{figure*}[hbt]
	\centering
	\includegraphics[width=0.9\textwidth]{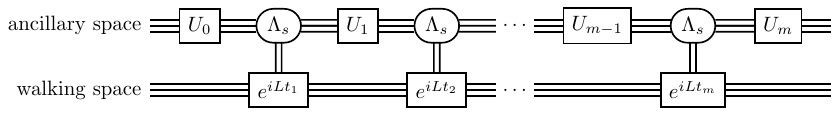}
	\caption{\label{fig:QW_model} Illustration of a CIQW $W$ with $s$ ancillary qubits and $m$ intermittent steps.
    The dumbbell-shaped gates denote the controlled CTQW $\Lambda_s(e^{iLt_j})=\sum_{l} \ket{l}\bra{l}\otimes  e^{ilLt_j}$. }
\end{figure*}

Suppose a graph $G$ has the Laplacian matrix $L$.
The {\it controlled intermittent quantum walk} (CIQW) on $G$ has the state space $\mathcal{K}\otimes\mathcal{H}$, where $\mathcal{K}$ consisting of $s$ qubits denotes the ancillary space that stores the control signal, and  $\mathcal{H} = \mathrm{span} \{ \ket{v} : v\in V(G) \}$  spanned by all vertices of graph $G$ is the walking space.
A CIQW (with $m$ intermittent steps), denoted by $W$, is defined as follows:
\begin{equation}\label{eq:CIQW_evolution}
    W := \prod_{j=1}^{m} \left[ (U_{j} \otimes I)\cdot \Lambda_s(e^{iLt_j}) \right]
    \cdot (U_0 \otimes I),
\end{equation}
where $U_j$ for $j\in\{0,1,\dots,m\}$ are unitary operators changing the control signal in the ancillary space $\mathcal{K}$,
and %$\Lambda_k(e^{iLt_j}): \ket{l} \ket{\psi} \mapsto \ket{l} \otimes e^{iLlt_j}\ket{\psi}$ 
$\Lambda_s(e^{iLt_j})=\sum_{l} \ket{l}\bra{l}\otimes  e^{ilLt_j}$ denotes the controlled unitary transformation that applies $(e^{iLt_j})^l$ to $\mathcal{H}$, controlled by  $l\in\{0,1,\dots,2^s-1\}$ in the ancillary space $\mathcal{K}$.
We use the notation $\prod_{j=1}^{m} A_j := A_m A_{m-1} \cdots A_1$ (rather than $A_1 A_2 \cdots A_m$), since $\mathcal{H}$ consists of column vectors and the rightmost $A_1$ is applied first.
An illustration of the CIQW $W$ is shown in Fig.~\ref{fig:QW_model}.

% state evolution Formula, what parameters we care, Hamiltonian simulation linear in time, overview of the algorithm, value of ancillary qubits number k

% \note{in the algorithms proposed here we only need the controlled-$e^{iLt}$ rather than $\Lambda_k(e^{iLt_j})$, which needs $k$ controlled-$e^{iLt}$ to implement}
% \note{$e^{i\pi \Pi_M}$? The model does not involve the oracle? Yes, here we just define quantum walk, and only when later using the model to tackle spatial search, we involve the oracle}
% The unitary transformation $\Lambda_k(e^{iLt_j})$ can be implemented with $k$ successive calls to the controlled CTQW c-$e^{iLt_j\cdot 2^j} = \ket{0}\bra{0}\otimes I +\ket{1}\bra{1}\otimes e^{iLt_j \cdot 2^j}$ for $j\in\{0,1,\dots,k-1\}$ (see for example the box in Fig.~\ref{fig:phase_estimation}), and the total evolution time is $T_j := t_j \cdot (2^0 +2^1 +\cdots + 2^{k-1}) = (2^k-1)t_j$.
As shown in Fig.~\ref{fig:QW_model}, a CIQW  performs a controlled CTQW $\Lambda_s(e^{iLt_1})$ for a period of time under a certain control signal generated by $U_0$, then adjusts the control signal by $U_1$ and performs a controlled CTQW $\Lambda_s(e^{iLt_2})$ for another period of time under the new control signal, repeating the operation in this way.
%We assume the unitary operators $U_j$ for $j\in\{0,1,\dots,m\}$ can be implemented efficiently, which is indeed the case in our algorithms,
%The main cost of implementing CIQW $W$ comes from the implementaion of $\Lambda_k(e^{iLt_j})$.
The main cost of implementing CIQW $W$ comes from the implementation of $e^{iLt_j}$. According to results on Hamiltonian simulation (Appendix~\ref{subsec:simulation}), %~\cite{berry2012black,Hamiltonian_2017,Low2019hamiltonian}, 
the circuit cost of simulating $e^{iHt}$ is linear in the evolution time $t$.
\footnote{The unitary transformation $\Lambda_s(e^{iLt_j})$ can be implemented with $k$ successive calls to the controlled CTQW c-$e^{iLt_j 2^j} = \ket{0}\bra{0}\otimes I +\ket{1}\bra{1}\otimes e^{iLt_j 2^j}$ for $j\in\{0,1,\dots,k-1\}$,
% (see for example the box in Fig.~\ref{fig:phase_estimation}),
and thus the total evolution time is $T_j := t_j \cdot (2^0 +2^1 +\cdots + 2^{s-1}) = (2^s-1)t_j$.}
% In our algorithms proposed later, will give the exact1 expression of $T$.

\section{Deterministic search on Laplacian integral graphs}\label{sec:exact_search}

\subsection{Algorithm overview}\label{subsec:overview}

\begin{figure}[hbt]
	\centering	\includegraphics[width=0.5\textwidth]{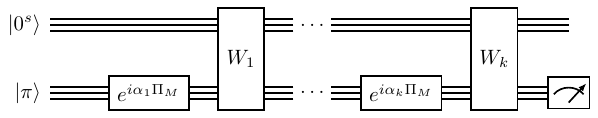}
	\caption{\label{fig:QWS} Illustration of our CIQW-based algorithm. }
\end{figure}

Given a graph with Laplacian matrix $L$ and a marked set $M$, our CIQW-based algorithm starts from the initial state $\ket{\pi}$, the uniform superposition of all the vertices, and then performs the CIQW and the oracle $e^{i\pi \Pi_M}$ alternately.
The oracle $e^{i\alpha \Pi_M}$ multiplies a relative phase of $e^{i\alpha}$ to vertices $\ket{v}$ in $M$.
An illustration of our algorithm is shown in Fig.~\ref{fig:QWS}.

The main idea is to mimic Grover's search algorithms, or more precisely, to implement the generalized Grover's iteration $G(\alpha,\beta) = e^{i\beta\ket{\pi}\bra{\pi}} \cdot e^{i\alpha \Pi_M}$.
As the oracle $e^{i\alpha \Pi_M}$ is already provided by Eq.~\eqref{eq:general_oracle_def}, the key is to implement exactly the phase shift $e^{i\beta\ket{\pi}\bra{\pi}}$ around the initial state $\ket{\pi}$ using CIQW.

For Laplacian integral graphs, we will set $k= O(\log(\lambda_{N}))$ in Fig.~\ref{fig:QW_model}.
This allows us to use phase estimation on $e^{iLt}$ to distinguish exactly the zero eigenvalue of $L$ from others, so as to construct a perfect phase shift $e^{i\beta\ket{\pi}\bra{\pi}}$ of $\ket{\pi}$.
As a result, we can use any of the deterministic quantum search schemes ~\cite{amplitude_amplification,arbi_phase,Long,FXR} to find a marked vertex with certainty, provided the number of marked vertices is known in advance.
The details are shown below.

\subsection{Proof of Theorem~\ref{thm:known}}\label{subsec:proof_thm}

In this subsection, we prove Theorem~\ref{thm:known}.
% That is, we will show that a marked vertex on the graph can be found with certainty in time $O(1/\sqrt{\varepsilon})$ if the spectrum of a Laplacian $L$ consists entirely of integers.
% Note that compared to the general case, the total evolution time now depends only on the proportion $\varepsilon$ of marked vertices, which we require to be known in advance.
Our CIQW-based deterministic spatial search algorithm for Laplacian integral graph is shown in Algorithm~\ref{alg:CIQW}.

\begin{algorithm}
\caption{Deterministic quantum search on Laplacian integral graphs based on CIQW}
\label{alg:CIQW}
% \begin{enumerate}
    1. Prepare the uniform superposition of all vertices $\ket{\pi} =\frac{1}{\sqrt{N}} \sum_{v\in V} \ket{v}$.
    % as the initial state, which is also the unique eigenstate of the graph Laplacian $L$ with eigenvalue $0$.
    
    2. Repeat the following two steps for $k =O(1/\sqrt{\epsilon})$ times, with parameters $\alpha, \beta, k$ set by Lemma~\ref{lem:long}.
    % , to transform from $\ket{\pi}$ to $\ket{M}$ (the uniform superposition of all marked vertices):

    \begin{enumerate}
    \item[(a)] Perform a CIQW (Fig.~\ref{fig:QW_model}) with $m=2$ intermittent steps and $s = \lceil \log(\lambda_N) \rceil$ ancillary qubits to implement exactly the relative phase shift $e^{i\beta \ket{\pi}\bra{\pi}}$ as shown in Fig.~\ref{fig:rotation_pi}.
    
    \item[(b)] Invoke the oracle $e^{i\alpha \Pi_M}$ shown in Fig.~\ref{fig:oracle}.
    %(Eq.~\eqref{eq:general_oracle_def}).
    \end{enumerate}

    3. Measure the walking space.
    % to obtain a marked vertex with certainty.
% \end{enumerate}
\end{algorithm}

\begin{figure*}[htb]
\includegraphics[width=0.8\textwidth]{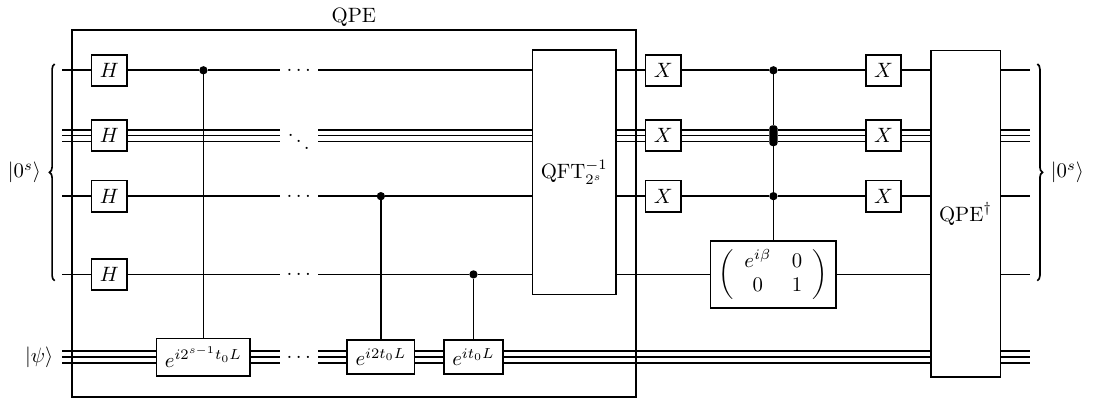}% Here is how to import EPS art
\caption{\label{fig:rotation_pi} Quantum circuit implementation of $e^{i\beta\ket{\pi}\bra{\pi} }$, where $s=\lceil\log(\lambda_N+1)\rceil$ and $t_0 = \pi/2^{s+1}$. }
\end{figure*}

We first show that the quantum circuit shown in Fig.~\ref{fig:rotation_pi} implements exactly the relative phase shift of the state $\ket{\pi}$, i.e.
\begin{equation}
    e^{i\beta \ket{\pi}\bra{\pi}} = I-(1-e^{i\beta})\ket{\pi}\bra{\pi},
\end{equation}
by applying quantum phase estimation (QPE)~\cite{phase_estimation} to $e^{iLt}$.

Suppose the eigenvalues of the graph Laplacian $L$ are
$0 =\lambda_1 < \lambda_2 \leq \cdots \leq \lambda_{N} \leq N$, with corresponding eigenvectors $\ket{\psi_0} =\ket{\pi}, \ket{\psi_1},\dots,\ket{\psi_{N-1}}$.
Let $s$ be the smallest integer such that $2^s > \lambda_{N}$ (i.e. $s=\lceil \log(\lambda_N+1) \rceil$) and set
\begin{equation}
    t_0 := \frac{\pi}{2^{s-1}}.
\end{equation}
Then the eigenvalues of $U := e^{it_0L}$ are $\{2\pi \frac{\lambda_k}{2^s}:k=1,\cdots,N\}$, where $\lambda_k \in \{0,1,\cdots,2^s-1\}$.
Note that $\ket{\psi_k}$ is an eigenvector of $U$ with corresponding eigenphase $\theta = 2\pi \frac{\lambda_k}{2^s}$, i.e. $U\ket{\psi_k} = e^{2\pi i \frac{\lambda_k}{2^s}} \ket{\psi_k}$.
A quick review of QPE (as shown by the box in Fig.~\ref{fig:rotation_pi}) is presented below.
After applying $H^{\otimes s} \otimes I$ to $\ket{0^s}\ket{\psi_k}$, we obtain $\frac{1}{\sqrt{2^s}} \sum_{j=0}^{2^s-1} \ket{j} \ket{\psi_k}$.
By the binary representation $j = j_{s-1}2^{s-1} +\cdots +j_{1}2^{1} +j_02^0$, where $j_i \in \{0,1\}$ for $i\in\{0,1,\cdots,s-1\}$, we know that the controlled-$U^{2^i}$ operations implement the transformation $\ket{j}\ket{\psi_k} \mapsto \ket{j} U^j \ket{\psi_k}$, and thus we obtain $\frac{1}{\sqrt{2^s}} \sum_{j=0}^{2^s-1} e^{2\pi i \frac{j\lambda_k}{2^s}} \ket{j} \ket{\psi_k}$.
Note that the state in the first $s$ qubits is exactly $\mathrm{QFT}_{2^s} \ket{\lambda_k}$.
Thus, by applying the inverse quantum Fourier transform $\mathrm{QFT}_{2^s}^{-1}$ to the first $s$ qubits, we obtain $\ket{\lambda_k}$.

Since the eigenvalues $\lambda_k$ for $k=1,2,\dots,N$ of the graph Laplacian $L$ have been obtained exactly, we can now add a relative phase shift of $e^{i\beta}$ to the eigenvector $\ket{\psi_0} =\ket{\pi}$ with eigenvalue $\lambda_1=0$, by applying a $(s-1)$-zero-controlled $e^{i\beta}\ket{0}\bra{0}+\ket{1}\bra{1}$ gate to the first $s$ qubits.
Finally, by reversing the phase estimation process to set the first $s$ qubits back to zero, we have successfully implemented $e^{i\beta \ket{\pi}\bra{\pi}}$.

The controlled CTQW c-$e^{iLt}$ is called for $2s = 2 \lceil \log(\lambda_{N}+1) \rceil < 2\lceil \log(N+1) \rceil $ times, and the total evolution time is
\begin{align}
    T &:= 2t_0(2^{s-1}+\cdots +2 +1)\\
    &= 4\pi (1-\frac{1}{2^s}) < 4\pi. \label{eq:t1_exact}
\end{align}

\begin{figure}[hbt]
	\includegraphics[width=0.35\textwidth]{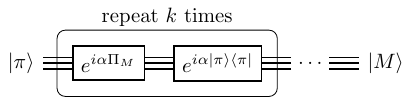}	\caption{\label{fig:exact_grover} Illustration of the deterministic Grover search by Lemma~\ref{lem:long}. }
\end{figure}

To achieve deterministic search on Laplacian integral graphs, we will use deterministic Grover search algorithms.
There are at least three schemes to achieve this~\cite{amplitude_amplification,arbi_phase,Long}.
% (see also \cite[Table 1]{FXR} for a summary).
But the following scheme due to Long~\cite{Long} is most concise, as it uses only one parameter $\alpha$ in the generalized Grover's iteration $G(\alpha,\beta)$.
An illustration of Long's deterministic Grover search is shown in Fig.~\ref{fig:exact_grover}.

\begin{lemma}[\cite{Long}]\label{lem:long}
    Denote by
    \begin{equation}
        G(\alpha,\beta) := e^{i\beta\ket{\pi}\bra{\pi}} \cdot e^{i\alpha \Pi_M}
    \end{equation}
    the generalized Grover's iteration,
    where $e^{i\beta\ket{\pi}\bra{\pi}}$ adds a relative phase shift of $e^{i\beta}$ to the initial state $\ket{\pi}$,
    and $e^{i\alpha \Pi_M}$ adds a relative phase shift of $e^{i\alpha}$ to all the marked elements.
    Suppose the proportion of marked elements $\varepsilon = \| \Pi_M \ket{\pi} \|^2$ is known in advance.
    Let $\alpha =\beta= 2\arcsin\left( {\sin(\frac{\pi}{4k+2})}/{\sqrt{\varepsilon}} \right) \in(0,\pi)$ and the iteration number $k > k_\mathrm{opt}$, where $k_\mathrm{opt} = \frac{\pi}{4\arcsin\sqrt{\varepsilon}} -\frac{1}{2}$.
    Then a marked element can be found exactly:
    \begin{equation}
        \left| \bra{M} G(\alpha,\alpha)^k \ket{\pi} \right| =1,
    \end{equation}
    where $\ket{M} := \Pi_M \ket{\pi} / \sqrt{\epsilon}$.
\end{lemma}

Since $e^{i\beta\ket{\pi}\bra{\pi}}$ and $e^{i\alpha \Pi_M}$ are implemented by Step~2.(a) and Step~2.(b) of Algorithm~\ref{alg:CIQW} respectively, Step~3 will obtain a marked vertex with certainty by Lemma~\ref{lem:long}.
This proves the correctness of Algorithm~\ref{alg:CIQW}.

We now calculate the total cost as follows.
Recall that the operator $e^{i\beta\ket{\pi}\bra{\pi}}$ can be implemented using $2 s =O(\log(N))$ controlled CTQW c-$e^{iLt}$ and with evolution time $T<4\pi$.
We set the iteration number of $G(\alpha,\beta)$ to be $k = \lceil k_\mathrm{opt} \rceil = O(1/\sqrt{\varepsilon})$.
% Note that the evolution time $\alpha$ in the marked vertex shift operator $S_M(\alpha) = e^{i\alpha \Pi_M}$ is less than $\pi$.
In total, the controlled CTQW c-$e^{iLt}$ is called for
\begin{equation}\label{eq:calls_known}
    O\left(\frac{\log(N)}{\sqrt{\varepsilon}}\right)
\end{equation}
times, and the total evolution time $T'$ and the number $k$ of invocation to the general oracle operator $e^{i\alpha \Pi_M}$ are as follows:
\begin{equation}\label{eq:p_T_known}
    T' = k T = O\left(\frac{1}{\sqrt{\varepsilon}}\right), 
    \quad k = O\left(\frac{1}{\sqrt{\varepsilon}}\right).
\end{equation}
This proves Theorem~\ref{thm:known}.
% achieving quadratic speedup similar to Grover's search.

\subsection{Gate complexity}\label{subsec:gate_complexity}
% For a general graph with $N$ vertices, its Laplacian $L$ has eigenvalues as $0 =\lambda_1 < \lambda_2 \leq \cdots \leq \lambda_{N} \leq N$.

% We will show in Section~\ref{subsec:proof_thm} that the number of elementary gates used in our algorithm apart from implementing the CTQW $e^{iLt}$ is $O\left(\frac{\log^2(\lambda_N)}{\sqrt{\varepsilon}}\right)$ (Eq.~\eqref{eq:gate_others_known}).

% \subsubsection*{Gate complexity}
% The gate complexity of simulating the CTQW $e^{iLt}$ simply equals the total evolution time $T = O(\frac{1}{\sqrt{\varepsilon}})$ multiplied by $O(\lambda_N N^4 \log(N))$, according to Eq.~\eqref{eq:CTQW_time} in Corollary~\ref{coro:CTQW_time}.

We first show that the total number of elementary gates used in Algorithm~\ref{alg:CIQW} apart from the CTQW $e^{iLt}$ is $O(\frac{\log^2(\lambda_N)}{\sqrt{\varepsilon}})$. 
From Fig.~\ref{fig:rotation_pi}, the dominating cost of implementing $e^{i\beta\ket{\pi}\bra{\pi}}$ comes from the quantum Fourier transform $\mathrm{QFT}_{2^s}^{-1}$, which costs $O(s^2) = O(\log^2(\lambda_N))$ elementary gates at most, and the $(s-1)$-controlled gate can be decomposed into $O(s)$ elementary gates~\cite{nielsen_chuang_2010}.
Since $e^{i\beta\ket{\pi}\bra{\pi}}$ is called for $k = O(1/\sqrt{\varepsilon})$ times, the total number of elementary gates apart from implementing the CTQW $e^{iLt}$ is
\begin{equation}\label{eq:gate_others_known}
    O\left(\frac{\log^2(\lambda_N)}{\sqrt{\varepsilon}}\right).
\end{equation}
% where $\lambda_N$ is the largest eigenvalue of the graph Laplacian $L$, and $\varepsilon$ is the proportion of marked vertices.

% \begin{remark}\label{rem:known}
% Suppose the CTQW $e^{iLt}$ can be implemented exactly with gate complexity $C(t) = O(\mathrm{poly}\log(N))$ for any time $t$, then the total gate complexity is
% \begin{equation}
%     C = O\left(\frac{\mathrm{poly}\log(N)}{\sqrt{\varepsilon}}\right),
% \end{equation}
% since we have shown above Eq.~\eqref{eq:p_T_known} that the controlled CTQW c-$e^{iLt}$ is called for $O(\frac{\log(\lambda_{N})}{\sqrt{\varepsilon}}) = O(\frac{\log(N)}{\sqrt{\varepsilon}})$ times in total.
% \end{remark}

The gate complexity of our algorithm is dominated by the implementation of $e^{iLt}$.
A trivial way of implementing  $e^{iLt}$ is by using product formulas and decomposing $L$ into Pauli strings, and it costs $O(\lambda_N t N^4 \log(N))$ elementary gates as shown by Corollary~\ref{coro:CTQW_time} in Appendix~\ref{subsec:simulation}.
Substituting the total evolution time into the aforementioned formula, the gate complexity is $O\left( \frac{\lambda_N N^4 \log(N)}{\sqrt{\varepsilon}} \right)$ for Laplacian integral graphs.

When there is an efficient quantum circuit for CTQW on graphs, the gate complexity can be significantly reduced.
By efficient we mean the CTQW $e^{iLt}$ can be implemented with gate complexity $O(\mathrm{poly}(\log(N)))$ independent of $t$.
In this case, we are concerned with the number of calls to $e^{iLt}$ instead of the total evolution time $T$.
We have shown that this number is equal to $O(\frac{\log(N)}{\sqrt{\varepsilon}})$ (Eq.~\eqref{eq:calls_known}).
Thus, we have the following result:
\begin{theorem}\label{rem:gate}
    When the CTQW $e^{iLt}$ on a graph $G$ with Laplacian $L$ can be implemented with gate complexity $O(\mathrm{poly}\log(N))$ independent of $t$,
    the algorithm in Theorem~\ref{thm:known} has gate complexity $O\left(\frac{\mathrm{poly}(\log(N))}{\sqrt{\varepsilon}}\right)$.
\end{theorem}

Efficient quantum circuit implementations for CTQW have been demonstrated across various graph classes, such as commuting graphs and Cartesian products of graphs~\cite{loke2017efficient1}.
Notably, certain graph classes, including complete bipartite graphs, complete graphs~\cite{Xu_2009,Childs2010} and star graphs~\cite{Xu_2009,alternating_PRL}, can be efficiently implemented using diagonalization techniques. Additionally, a specific set of circulant graphs can be efficiently implemented, leveraging the quantum Fourier transform~\cite{qiang2016efficient}. For circulant graphs with $2^n$ vertices possessing $O(\text{poly}(n))$ non-zero eigenvalues or efficiently characterizable distinct eigenvalues (e.g., cycle graphs, Möbius ladder graphs), efficient quantum circuits for CTQW can be constructed~\cite{qiang2016efficient}.
Ref.~\cite{chen2024CTQW} discussed how to implement CTQW on general graphs based on graph decomposition.

\section{Conclusion and discussion}\label{sec:summary}
In this paper, we have addressed the deterministic spatial search problem,  and proposed a generic algorithm based on the recently proposed controlled intermittent quantum walk (CIQW) model.
Our algorithm can find with certainty a marked vertex on any Laplacian integral graphs in total evolution time $O(\frac{1}{\sqrt{\varepsilon}})$ and with query complexity $O(\frac{1}{\sqrt{\varepsilon}})$ as long as the proportion of marked vertices $\varepsilon$ is known in advance.
This improves previous result~\cite{universal} based on alternating quantum walks, as their algorithm has worse complexities on certain graphs, only works for a single marked vertex, and requires additionally the graph to be vertex transitive.
We also give an analysis of the gate complexities of our algorithm, which are dominated by the quantum circuit implementation for the CTQW $e^{iLt}$.
It will be interesting to discover more classes of graphs on which the CTQW $e^{iLt}$ has efficient quantum circuit implementation.

\begin{acknowledgments}
This work was supported by the National Key Research and Development Program of China (Grant No.2024YFB4504004), the National Natural Science Foundation of China (Grant No. 92465202, 62272492, 12447107),  the Guangdong Provincial Quantum Science Strategic Initiative (Grant No. GDZX2303007, GDZX2403001), the Guangzhou Science and Technology Program (Grant No. 2024A04J4892).
\end{acknowledgments}

\appendix

\section{Laplacian integral graphs}\label{subsec:integral}
A graph whose spectrum of its Laplacian consists entirely of integers is called a \textit{Laplacian integral} graph~\footnote{Note that there are no Laplacian `rational' graphs. This is because if the eigenvalue of a graph is rational, then it is necessarily an integer, see \url{https://math.stackexchange.com/questions/2936232} and \url{https://math.stackexchange.com/questions/3123348}.}.
In general, the problem of characterizing Laplacian integral graphs seems to be very difficult, and there are many results on the finding of particular classes of Laplacian integral graphs~\cite{Laplacian_maximal,survey_integral,Laplacian_distinct,Laplacian_Sab,Laplacian_indecomposable,Laplacian_pf}.
We list below some examples of Laplacian integral graphs, all of which allows for deterministic spatial search by Theorem~\ref{thm:known}.
\begin{enumerate}
    \item Complete graph $K_n$.
    This graph has $n$ vertices, and each vertex is connected to the other $n-1$ vertices.
    The Laplacian eigenvalues are $0$ and $n$ with corresponding multiplicities $1$ and $(n-1)$.

    \item Johnson graph $J(n,k)$~\cite{regular}.
    The vertex set consists of all subset $S$ of $[n]:=\{1,2,\dots,n\}$ with size $k$, and two vertices $S,S'$ are adjacent if and only if $|S\cap S'|=k-1$.
    The Laplacian eigenvalues are $i(n+1-i)$ for $i\in\{0,1,\dots,\min(k,n-k)\}$, with corresponding multiplicities $\binom{n}{i} -\binom{n}{i-1}$.
    We define $\binom{n}{-1}=0$.

    \item Kneser graph $K(n,k)$~\cite{regular}.
    The vertex set consists of all subset $S$ of $[n]:=\{1,2,\dots,n\}$ with size $k<n/2$, and two vertices $S,S'$ are adjacent if and only if $S\cap S'=\emptyset$.
    The Laplacian eigenvalues are $\binom{n-k}{k} -(-1)^i\binom{n-k-i}{k-i}$ for $i\in\{0,1,\dots,k\}$, with corresponding multiplicities $\binom{n}{i} -\binom{n}{i-1}$.

    \item Hamming graph $H(d,q)$~\cite{regular}.
    The vertex set consists of all $d$-tuples $(x_1,\dots,x_d)\in[q]^d$, and two vertices $x,x'$ are adjacent if and only if they differ in exactly one coordinate.
    The Laplacian eigenvalues are $qi$ for $i\in\{0,1,\dots,d\}$, with corresponding multiplicities $\binom{d}{i}(q-i)^i$.
    Note that the hypercube is $H(n,2)$.

    \item Grassmann graph $G_q(n,k)$~\cite{Brouwer1989}.
    The vertex set consists of all $k$-dimensional subspaces of $\mathbb{F}_q^n$, the $n$-dimensional space over the finite field $\mathbb{F}_q$, and two vertices $A,B$ are adjacent if and only if $\dim(A\cap B) =k-1$.
    There are $\binom{n}{k}_q := \frac{(q^n-1)\cdots(q^{n-m+1}-1)}{(q^m-1)\cdots(q-1)}$ vertices in total, and all the vertices have the same degree $q[k]_q[n-k]_q$, where $[k]_q := \frac{q^k-1}{q-1}$.
    The Laplacian eigenvalues are $q [k]_q [n-k]_q -q^{i+1} [k-i]_q [n-k-i]_q +[i]_q$ for $i\in\{0,1,\dots,\min(k,n-k)\}$, with corresponding multiplicities $\binom{n}{i} -\binom{n}{i-1}$.

    \item Rook graph $R(m,n)$~\cite{alternating_QST}.
    It is the Cartesian product $K_m \square K_n$ of the $m$-vertex complete graph $K_m$ and the $n$-vertex complete graph $K_n$.
    Specifically, there are $mn$ vertices in total: $V = \{(u,v): u\in V(K_m), v\in V(K_n)\}$, and two vertices $(u,v)$ and $(u',v')$ are adjacent if and only if either $u=u'$ and $(v,v')\in E(K_n)$; or $v=v'$ and $(u,u')\in E(K_m)$.
    All the vertices have the same degree $(m+n-2)$.
    The Laplacian eigenvalues are $0, n, m, (n+m)$, with corresponding multiplicities $1, (n-1), (m-1),(m-1)(n-1)$.
    Note that the $2n$-vertex complete identity interdependent network (CIIN) is equivalent to the Rook graph $R(2,n)$~\cite{alternating_PRA}.

    \item Complete-square graph $K_n \square Q_2$~\cite{alternating_QST}.
    It is the Cartesian product of an $n$-vertex complete graph $K_n$ with a square graph $Q_2$.
    All the vertices have the same degree $(n+1)$.
    The Laplacian eigenvalues are $0,2,4,n,(n+2),(n+4)$ with corresponding multiplicities $1,2,1,(n-1),2(n-1),(n-1)$.
    
    \item Cocktail-party graph $CP(n)$~\cite{survey_integral}.
    The vertex set is $V = \{(u,b):u\in[n],b=0,1\}$, and each vertex $(u,b)$ is connected to all the other $(2n-2)$ vertices except the vertex $(u,1-b)$.
    The Laplacian eigenvalues are $0,(2n-2),2n$ with corresponding multiplicities $1,n,(n-1)$.

    \item Complete $k$-partite graph $K_{n/k,n/k,\dots,n/k}$~\cite{survey_integral}.
    The $n$ vertices are equally divided into $k$ classes, and each vertex $v$ is connected to the other $(n-n/k)$ vertices that are not in the same class as $v$.
    The Laplacian eigenvalues are $0, (n-n/k), n$ with corresponding multiplicities $1,(n-k),(k-1)$.

    \item Star graph $K_{1,n}$~\cite{survey_integral}. It is a tree with one node having degree $n$ and the other $n$ vertices having degree $1$.
    The Laplacian eigenvalues are $0,1,(n+1)$ with corresponding multiplicities $1,(n-1),1$.    
\end{enumerate}

\section{Hamiltonian Simulation}\label{subsec:simulation}
% Suppose a Hamiltonian $H \in \mathbb{C}^{M\times M}$ has at most $D$ nonzero elements in each row or column, and is provided by the following two oracles:
% \begin{align}
% 	O_H \ket{j,k}\ket{z} &= \ket{j,k}\ket{z\oplus H_{jk}}, \\
% 	O_F \ket{j,l} &= \ket{j,f(j,l)},
% \end{align}
% where $j,k \in\{1,2\dots,M\}$, $l\in\{1,2,\dots,D\}$.
% The function $f(j,k)$ gives the row index of the $k$-th nonzero element in column $j$ (or the row index of any zero element when there are fewer than $k$ nonzero elements in column $j$).
% The matrix element $H_{jk}$ is represented by its real and imaginary parts written in binary, and $\oplus$ is the bitwise XOR of such representations.
% Let $\| H \|$ denotes the spectral norm of $H$, and $\| H \|_\mathrm{max} := \max_{j,k} |H_{j,k}|$.
% Then $e^{iHt}$ can be simulated with scaling linear in both $\|H\|t$ and $D$:
% \begin{lemma}[\cite{berry2012black}]
%     For a given Hamiltonian $H$, let $\Lambda \geq \| H \|$ and $\Lambda_\mathrm{max} \geq \| H \|_\mathrm{max}$. Then $e^{iHt}$ can be simulated with error at most $\delta\in(0,1]$ using
%     \begin{equation}
% 	   O \left( \frac{\Lambda t}{\sqrt{\delta}} + D\Lambda_\mathrm{max}t +1 \right)
%     \end{equation}
%     queries to $O_H$ and $O_F$.
% \end{lemma}

Product formulas method~\cite{lloyd1996universal} for the problem of Hamiltonian simulation is commonly used to simulate the evolution $e^{-iHt}$ when Hamiltonian has the form $H=\sum_{j=1}^m H_j$, where each $H_j$ can be efficiently simulated for arbitrary evolution time $t'$.
Then the evolution $e^{-iHt}$ can be approximated by a product of exponentials $e^{-iH_jt'}$ within error $\epsilon$.
The upper bound of the number of exponentials required is given in Ref.~\cite{berry2007efficient}, as shown in the following Theorem~\ref{theorem:pf}.
% \note{it seems the theorem is not used.}
\begin{theorem}[\cite{berry2007efficient}, Theorem~1]\label{theorem:pf}
    In order to simulate the Hamiltonian of the form $H=\sum_{j=1}^m H_j$ for time $t$ by a product of exponentials $e^{-iH_jt'}$ within error $\epsilon$, the number of exponentials $N_{\exp}$ is bounded by
  \begin{equation}
        N_{\exp} \le 2m^25^{2k} \|H\|t(m\|H\|t/\epsilon)^{1/2k},
   \end{equation}
   for $\epsilon \le 1 \le 2m5^{k-1} \|H\| t$, where $\| H \|$ denotes the spectral norm of $H$, and $k$ is an arbitrary positive integer.
\end{theorem}

The total gate count of quantum circuit $\tilde{U}$ is upper bounded by $N_{\exp} \times \max_j C(H_j)$, where $C(H_j)$ is the cost required to implement exponential $e^{-iH_jt'}$.
In the worst case, every Hamiltonian $H \in \mathbb{C}^{N\times N}$ can be decomposed into $m=N^2$ Pauli strings $H_j \in \{ I,X,Y,Z\}^{\otimes n}$, assuming for simplicity $N=2^n$.
Since each $e^{-iH_jt'}$ can be simulated with $C(H_j) =O(n) = O(\log(N))$ elementary gates, and $k$ is an arbitrary positive integer, we have the following corollary of Theorem~\ref{theorem:pf}.

\begin{corollary}\label{coro:CTQW_time}
    In order to simulate the CTQW $e^{iLt}$ for time $t$ within constant error, the number of elementary gates is bounded by
    \begin{equation}\label{eq:CTQW_time}
	   C(e^{iLt}) = O(\lambda_N t N^4 \log(N)),
    \end{equation}
    where $\lambda_N$ is the largest eigenvalue of the graph Laplacian $L$.
\end{corollary}

% Note that for certain graphs, including complete bipartite graphs, complete graphs~\cite{loke2017efficient1}, glued trees~\cite{CCD03} and star graphs~\cite{alternating_PRL}, the CTQW on them can be simulated more efficiently using diagonalization techniques.

% \nocite{*}

\bibliography{apssamp}% Produces the bibliography via BibTeX.

\end{document}